%Paper: hep-th/9304002
%From: devega@lpthe.jussieu.fr (Hector DE VEGA)
%Date: Thu, 1 Apr 93 17:27:30 +0200
%Date (revised): Fri, 2 Apr 93 12:02:00 +0200
%Date (revised): Tue, 6 Apr 93 15:28:26 +0200
%Date (revised): Sat, 17 Apr 93 16:14:55 +0200

\input phyzzx
%\input myp

%%%%%%%%%%%%%%%%%%%%%%%%%%%%%%%%%%%%%%%%%%%%%%%%%%%%%%%%%%%%%%%%%%
\def\AP#1{{\sl Ann.\ Phys.\ (N.Y.) {\bf #1}}}

\def\IMPA#1{{\sl Int. J. Mod. Phys. {\bf A#1}}}

\def\MPL#1{{\sl Mod.\ Phys.\ Lett. \ {\bf #1}}}

\def\NPB#1{{\sl Nucl.\ Phys.\ {\bf B#1}}}

\def\PLB#1{{\sl Phys.\ Lett.\ {\bf #1B}}}

\def\PRD#1{{\sl Phys.\ Rev.\   {\bf D #1}}}
\def\PRL#1{{\sl Phys.\ Rev.\ Lett.\ {\bf #1}}}

\def\TMP#1{{\sl Theor.\ Math.\ Phys.\ {\bf #1}}}

%############################################################

%##############################################################
\def\nxl{\hfill\break}
%##############################################################

                            % A

                            % C

                            % F
                            % G
                            % H
                            % I
                            % L
                            % M
                            % N
                            % P
                            % O
                            % T
                            % U
                            % V
                            % Z
%##############################################################

\def\a{\alpha}

\def\b{\beta}
\def\g{\gamma}

\def\l{\lambda}

\def\s{\sigma}

\def\t{\theta}

\def\Th{\Theta}
%##############################################################

\def\o{\over}

                          % overline
                      % dm
                     % udm
                % cotanh
                  % acosh
                            % vector x
                            % vector y
                            % vector q
                            % vector p
%##############################################################
\def\bold#1{\setbox0=\hbox{$#1$}%                %Truccaccio per boldface
     \kern-.025em\copy0\kern-\wd0
     \kern.05em\copy0\kern-\wd0
     \kern-.025em\raise.0433em\box0 }
\def\lowmp{\lower.11em\hbox{${\scriptstyle\mp}$}}

                % abs
            % set
   % Vacuum Expect. Value
\def\frac#1#2{{\textstyle{
 #1 \over #2 }}}                            % fraction

   % blank spaces

                    % traccia anche sul colore
                  % argomento di un numero complesso
                   % slash operator
                     % real numbers
                     % natural numbers
                       % integers
                    % complex
\def\1{{\rm 1 \!\!\, l}}                        % 1I
 % commutator
 % anti-commutator
 % scal (,)
               % bra
               % ket
%
% derivate parziali: accetta anche \ e {} come argomento

%
%
% prodotto scalare tra un bra ed un ket

                                 % daga
                             % Q di BRS
%%%%%%%%%%%%%%%%%%%%%%%%%%%%%%%%%%%%%%%%%%%%%%%%%%%%%%%%%%%%%%%%%%%

\hyphenation{Di-par-ti-men-to}
\hyphenation{na-me-ly}
\hyphenation{al-go-ri-thm}
\hyphenation{pre-ci-sion}
\hyphenation{cal-cu-la-ted}

%

%

%---------------------------------------------------------------
\Pubnum={$\rm PAR\; LPTHE\; 93/14
         \qquad {\rm March \; 1993}$}
\date={}
\titlepage
%----------------------------------------------------------------
\title{{\bf THE TWO-DIMENSIONAL
STRINGY BLACK-HOLE:
A NEW APPROACH AND A PATHOLOGY.}}
\author{ H.J. de Vega }
\address{ Laboratoire de Physique Th\'eorique et Hautes Energies, Paris
     \foot{Laboratoire Associ\'e au CNRS UA 280 \nxl
      Postal address: \nxl
           L.P.T.H.E., Tour 16, $1^{\rm er}$ \'etage,
	Universit\'e Paris VI,\nxl
	4, Place Jussieu, 75252, Paris cedex 05, FRANCE }}
\author{  J. Ram\'irez Mittelbrun, M. Ram\'on Medrano }
\address{ Departamento de F\'isica Te\'orica, Madrid
\foot{Postal address: \nxl Facultad de Ciencias F\'isicas, Universidad
Complutense,  Ciudad Universtaria, \nxl E-28040, Madrid, ESPA\~NA.} }
\author{N. S\'anchez}
\address{Observatoire de Paris, Section de Meudon, Demirm
\foot{Laboratoire Associ\'e au CNRS UA 336, Observatoire de Meudon et
\'Ecole Normale Sup\'erieure.\nxl  Postal address: \nxl
DEMIRM, Observatoire de Paris. Section de Meudon, 92195 MEUDON
Principal Cedex, FRANCE.}}
%----------------------------------------------------------------
\endpage
%-----------------------------------------------------------------
\vfil
\abstract
The string propagation in the two-dimensional stringy black-hole is
investigated from a new  approach .
We completely solve the classical and quantum string dynamics
in the lorentzian and euclidean regimes. In the lorentzian case all
the physics reduces to a massless scalar particle described by a
Klein-Gordon type equation with a singular effective potential.
The scattering matrix is found and it reproduces the results obtained
by coset CFT techniques. It factorizes into two pieces : an elastic
coulombian amplitude and an absorption part. In both parts, an infinite
sequence of imaginary poles in the energy appear.
The generic features of string propagation in curved
D-dimensional backgrounds (string stretching, fall into
spacetime singularities) are analyzed in the present case.
A {\bf new} physical phenomenon specific to the present black-hole
is found  : the quantum renormalization of the speed of light.
We find  $c_{quantum} = \sqrt{{k\o{k-2}}}~c_{classical}$ ,
where $k$ is the integer in front of the WZW action.
 This feature is, however, a pathology. Only for $ k \to \infty $ the
pathology
disappears  (although  the conformal anomaly is present).
We analyze all the classical euclidean string solutions and exactly
compute the quantum partition function. No critical Hagedorn
temperature appears here.

 \endpage
\REF\bars{I. Bars and D. Nemeschansky, \NPB{348}, 89 (1991).\nxl
G. Mandal and A. M. Sengupta and S. R. Wadia, \MPL{A6}, 1685(1991).}
\REF\rabi{K. Bardacki, A. Forge and E. Rabinovici, \NPB{344},344
(1990).\nxl
S. Elitzur, A. Forge and E. Rabinovici, \NPB{359}, 581 (1991).}
\REF\Ed{E. Witten, \PRD{44} , 314 (1991).}
\REF\DVV{R. Dijkgraaf, H. Verlinde and E. Verlinde, \NPB{371}, 269 (1992).}
\REF\otros{I. Bars and K. Sfetsos, \MPL{A7}, 1091 (1992). \nxl
I. Bars in ``String Quantum Gravity and Physics at the Planck Energy Scale''
\nxl Erice School, N. S\'anchez Editor, World Sci. Publ. 1993. \nxl
M. Crescimanno, \MPL{A7}, 489 (1992). \nxl
E. S. Fradkin and V. Ya. Linetsky, \PLB{277}, 73 (1992). \nxl
J H Horne and G T Horowitz, \NPB{368} , 444 (1992).\nxl
C. J. Ahn et al., \PRD{47}, 1699 (1993).}
\REF\arka{A. A. Tseytlin, Imperial College preprint/TP/92-93/10}
\REF\nos{H. J. de Vega and N. S\'anchez,  \PLB{197}, 320 (1987).}
 \REF\negro{ H. J. de Vega and N. S\'anchez,
        \NPB{309}, 552 and 577(1988).}
\REF\erice{ H. J. de Vega, in ``String Quantum Gravity and
Physics at the Planck Energy Scale''
Erice School, N. S\'anchez Editor, World Sci. Publ. 1993,\nxl
 N. S\'anchez, ibidem.}
\REF\ondch{H. J. de Vega and N. S\'anchez, \NPB{317}, 706 and 731 (1989).}
\REF\prim{ H. J. de Vega and N. S\'anchez, \NPB{299}, 818(1988).}
\REF\costa{M. Costa
 and H. J. de Vega, \AP{211}, 223 and 235 (1991).}
 \REF\ondpl{H. J. de Vega and N. S\'anchez, \PRD{45} , 2783 (1992). \nxl
 H. J. de Vega, M. Ram\'on Medrano and N. S\'anchez,   LPTHE  Paris
  preprint  92-13 and Demirm Meudon preprint, to appear in Class. and
Quantum Grav.}
 \REF\desi{H. J. de Vega and N. S\'anchez,  LPTHE  Paris  preprint
  92-31 and Demirm Meudon preprint, to appear in \PRD{},\nxl
 H. J. de Vega, A. V. Mikhailov and N. S\'anchez,
 LPTHE  Paris  preprint 92-32 and Demirm Meudon preprint,
 to appear in \TMP{}, special volume in the memory of
 M. C. Polivanov. }
 \REF\vene{N. S\'anchez and G. Veneziano, \NPB{333}, 253 (1990), \nxl
 M. Gasperini, N.S\'anchez and G. Veneziano, \nxl
 \IMPA{6}, 3853 (1991) and \NPB{364}, 365 (1991).}
\REF\frolov{ V. P. Frolov and N. S\'anchez, \NPB{349}, 815 (1991).}
\REF\conico{H. J. de Vega and N. S\'anchez, \PRD{42} , 3969 (1990) and \nxl
 H. J. de Vega, M. Ram\'on Medrano and N. S\'anchez, \NPB{374}, 405
(1992).}
 \REF\plb{H. J. de Vega and N. S\'anchez, \PLB {244}, 215(1990), \nxl
    \PRL{65} (C), 1517(1990) and  \IMPA{7}, 3043 (1992).}
 \REF\trans{ H. J. de Vega, M. Ram\'on Medrano and N. S\'anchez,
 \NPB{351}, 277 (1991),\nxl \NPB{374}, 425 (1992) and  \PLB{285},206(1992) .}
\REF\carlos{C.O.Loust\'o and N. S\'anchez, \NPB{355}, 231 (1991),
\NPB{383}, 377 (1992) and \PRD{46}, 4520 (1992).}
\REF\agune{C.O.Loust\'o and N. S\'anchez, to appear in \PRD{}}
\REF\volodia{S L Lukyanov and V A Fateev, \nxl Soviet Scientific Reviews,
sect A, vol. 15, part 2, p. 117 .}
\medskip
\REF\ns{ N. S\'anchez, \PRD{24}, 2100 (1981), \nxl N. S\'anchez and
B. F. Whiting, , \PRD{34}, 1056 (1986).}

\vskip 1cm

\chapter{{\bf Contextual Background}}

The space-time metric associated to the $SL(2,R)/U(1)$ coset model (and
its different versions), and interpreted as a two dimensional black
hole [\bars ,\rabi ,\Ed ] , arised enormous attention recently [\DVV
,\otros ,\arka ].

In this paper, we look at this problem from our string gravity point of
view, that we started and developped (see refs.[\nos -\agune ]) for instance),
independently and before the interest on two dimensional black holes
flourished.

The two dimensional black hole model is interesting in the sense that
it is an exact solution of the renormalization group equations of string
theory. As it is known, these equations define the backgrounds in which
strings propagate consistently.

Although two dimensional models have many attractive tractable aspects
and can be used to test and get insights on particular features, $D=2$
is not for string theory neither for gravity the most physically appealing
dimension.

Strings in  two dimensional black hole backgrounds have been extensively
treated
in the literature in the context of the conformal Field Theory (CFT)
techniques. (See ref. [\DVV ]  for a complete description). In this paper, we
present a different view and a different approach to this problem which
yields in a simple and physical way the full exact quantum result;
giving  new insights with respect the already known CFT descriptions.

\chapter{{\bf Introduction}}

We study string propagation in the two-dimensional black hole background
of refs.[\bars -\Ed ]
$$
dS^2=k\left[ dr^2 - \tanh^2r \, dt^2 \right] ~,~
\Phi = \log (\cosh^2r) ~,~~ 0\leq r<\infty ,~ -\infty < t < \infty.
\eqn\fondo
$$

where $k$ is a positive integer and $\Phi$ is the dilaton field.
(Classically, the string propagates in this metric, and $\Phi$ is
decoupled from them). In Kruskal null type coordinates $(u ,v)$ the
metric eq.\fondo\ reads
$$
dS^2=k~{{du dv} \o {1 - uv}}~,~~0\leq u,v< \infty.
\eqn\krus
$$
We completely solve the string equations of motion and constraints in
the Lorentzian and on the Euclidean (i.e. $ t = i \Theta$ ) regimes. The
solutions fall into four types. In the Lorentzian regime in which there
are no compact dimensions in the space-time, the $\sigma$-dependence can
be completely gauged away, and all the solutions describe just the
geodesics of a massless point particle. The only physical degree of
freedom is the one associated to the center of mass of the string ; we
are left more with a point particle field theory rather than a string
theory.

In the Euclidean regime, the Schwarzchild imaginary time is periodic ,
($0 \leq \Th \leq 2\pi $), and then the $\sigma$-dependence
of the string solutions remains. The
solution must be also euclidean in the world sheet ($\sigma$ becames
purely imaginary and identified to the imaginary time). There are two
types of euclidean solutions : (i) a string winding $n$ times along the
$\Theta$ -direction in the "cigar" manifold $dS^2=k\left[ dr^2 +
\tanh^2r \, d\Th^2 \right]$
from $r=0$ to $r=\infty $,
and (ii) a string winding $n$ times along $\Theta$ in the "trumpet"
manifold $dS^2 = k\left[ds^2 + \coth^2s ~ d\Th^2\right]$, $0<s<\infty$.
 It must be noticed that these are {\it rigid strings} staying
around the manifolds without any oscillation. In $D=2, \; e^{i n \sigma}$
does not describes the string fluctuations, thus the name modes
("winding modes" or "momentum modes", as commonly refered (see for
example ref.[\DVV ]), are misleading.

For these solutions, the metric on the world-sheet is
$$\eqalign{
dS^2_{cigar} &= k\, {{n^2}\o{1+e^{2n\tau}}}~(d\tau^2+d\s^2) \cr
dS^2_{trumpet} &= k\, {{n^2}\o{1-e^{-2n\tau}}}~(d\tau^2+d\s^2)\cr}
\eqn\cigar
$$
The proper string length (between two points $\sigma$ and $\sigma + d
\sigma$ at fixed $\tau$) is
$$
\Delta S_{cigar} = \sqrt{k}\,{n\o{\sqrt{1+e^{2n\tau}}}}~ d\s ~\quad ,~\quad
\Delta S_{trumpet} = \sqrt{k}\, {n\o{\sqrt{1-e^{2n\tau}}}} ~d\s
\eqn\lonpro
$$

In $D=2$ the string world sheet and the physical space-time can be
identified ; the world sheet covers completely (generically, $n$ times)
the physical space. The trumpet manifold has a curvature singularity at
$s=0$. Near $s=0$, i.e.  $\tau \rightarrow 0$ , $\Delta S_{trumpet}$ blows up,
the string stretches infinitely. This typical feature of string
instability [\ondpl -\frolov ,\agune ] , just
corresponds here to the location of the string in
the open extremity of the trumpet. The euclidean black hole manifold is
non singular, and the string length ($\Delta S_{cigar}$), is always finite
in it, since the cigar radius is finite. We also discuss the
cosmological version of the black hole string solutions (in the euclidean
regime) ; in this case the string solutions are the same as the trumpet
solution.

We recall that in $D =2$ de Sitter space-time with Lorentzian
signature, the general solution is a string wound $n$ times around de
Sitter space (the circle $S ^1$) and evolving with it. For $\tau
\rightarrow 0$, $\Delta S_{de Sitter} \rightarrow \infty $ ; the string
expands (or contracts, $\Delta S_{de Sitter} \rightarrow 0$ ) with the
universe itself.

It is interesting to point out the relation between $\tau$ and the
physical time : $\tau$ interpolates between the Kruskal time
($\tau \rightarrow 0$) and the Schwarzschild time ($\tau \rightarrow \infty$).
This is like strings in cosmological backgrounds in which $\tau$
interpolates between the conformal time ($\tau \rightarrow 0$) and the
cosmic time ($\tau \rightarrow \infty$). (And like strings in gravitational
wave backgrounds). The logarithmic relation between the Schwarzchild
time and $\tau$ for $\tau \rightarrow 0$ is exactly like the cosmic
time and $\tau$ in de Sitter space.

In this two dimensional black hole, the string takes an infinite time
$\tau$ to approach the curvature singularity $u v = 1$, and then, never
reaches it. This is to be contrasted with the higher dimensional black
hole [\negro , \agune ]. For all $D>2$, the string falls down into the physical
singularity $u v = 1$ in a finite proper time $\tau$, and classically
as well as quantum mechanically, the string is trapped by the
singularity [\agune ].

We study the string quantization in this background (in the lorentzian
and in the euclidean regimes) and follow the spirit of ref.[\nos ], in
which one starts from the c.m. motion, and then takes the
$\sigma$-dependence as fluctuations around. Here, in the  lorentzian case,
there are no $\sigma$-fluctuations, and the quantization of the c.m.
Hamiltonian
$$
 \hat {\cal H}= - {1 \o {2k}}\left(
{{\partial^2} \o {\partial r^2}} + (\coth r + \tanh r )
{{\partial} \o {\partial r}}- \coth^2r ~ {{\partial^2} \o {\partial t^2}}
\right)
\eqn\hamcm
$$
yields the full and exact description of the system.

Eq.\hamcm\  is the same as the $L_0$ operator of ref.[\DVV] (hereafter
refered as DVV). They only differ by additive and multiplicative
constants, but as operators are identically the same). We solve the
eigenfunction
problem
$$
 \hat{\cal H} \Psi = {{\l} \o {2k}} \Psi
$$
describing the scattering of the (massless) tachyon field by the black
hole, which yields a Schr\"odinger type equation with an
energy-dependent potential
$$\eqalign{
V_{eff} =& -{{E^2+1/4}\o{\sinh^2r}}~ + ~ {1\o{4\cosh^2r}}
-E(E-1)-\l+1~ \cr
V_{eff} \buildrel{r\to0}\over =& -{\g \o{r^2}} + O(1) ~,
{}~ \g \equiv E^2+1/4  ~. \cr}
\eqn\vefec
$$
Since $\gamma \geq 1/4$, there exists absorption (fall into the event
horizon $r=0$) for all energy $E$. This is like the $D>2$ black holes.

The constant $\lambda$ is determined from the $r\rightarrow \infty$
behaviour, by requiring the tachyon to be massless
$$
\l-1 = (1-c^2) p^2
$$
$p$ being the momentum, $p = E/c$.

The eigenfunctions $\Psi$ are given in terms of hypergeometric
functions
$$
\Psi^{\pm}_{E,c}(r,t) = e^{-iEt} ~ (\sinh r)^{\mp iE}
(1+ {1\o c}),{1\o2}\mp {{iE} \o 2} (1- {1\o c});
1\mp iE \, ;-\sinh^2r %\right)
\eqn\hiper
$$
They exhibit the typical behaviour of the wave functions near the
black hole horizon
$$
\Psi^{\pm}_{E,c}(r,t)\buildrel{r\to0}\over = e^{-iE(t\pm\log r)} ~,
\eqn\hori
$$
$\Psi^+$ describing purely incoming particles at the future event
horizon. ($\Psi^-$ describes outcoming particles from the past
horizon). The tachyon field incident from spatial infinity is
partially absorbed and reflected by the black hole
$$
\Psi^{+}_{E,c}(r,t)\buildrel{r\to \infty}\over = e^{-ipct} \left[
e^{-ipr} + S(p,c)\,e^{ipr} \right]
\eqn\infi
$$
We find
$$
 S(p,c) = S_{coul}(p)~{\tilde S(p,c)}
$$
where
$$\eqalign{
 S_{coul}(p) &= 2^{-2ip}{{\Gamma(ip)}\o{\Gamma(-ip)}} =
2^{-2ip} e^{2i{\rm arg}\Gamma(ip)} \cr
{\tilde S(p,c)} &= \left[{{\Gamma\left({1\o2}-{{ip}\o2}(c+1)\right)}
\o {\Gamma\left({1\o2}-{{ip}\o2}(c-1)\right)}} \right]^2 \cr}
\eqn\maese
$$
$S_{coul}$ takes into account the large $r$ interaction and the purely
elastic scattering;  ${\tilde S}$ describes the genuine black hole features :
$$\eqalign{
| S_{coul}(p)| &= 1 \cr
| {\tilde S(p,c)}| &= \left[{{\cosh{{{\pi p}\o2}(c-1)}}\o
{\cosh{{{\pi p}\o2}(c+1)}}}\right]^2 ~<~1~,\cr}
\eqn\mods
$$
which describes the absorption by the black hole.

$S(p,c)$ exhibits an infinity sequence of imaginary poles at the
values $ip=n, (n=0,1,...)$, which are like the Coulombian-type poles,
and also an infinite sequence of imaginary poles at $\; ip(c+1) = 2n+1,\;
n=0,1...$ (in the ${\tilde S}$ part).

It must be stressed that $S(E,c)$ depends on two physical parameters
: $E$ and $c$ . For each energy $E$, we have a monoparametric family
of solutions depending on $c$, each $c$ yields a different
$S$-matrix. $c$ is a {\bf purely quantum mechanical parameter}, which is
not fixed by any special requirement, and can take any value ; $c$
accounts for a renormalization of the speed of light. Classically,
we choose our units such that $c=1$ (see eq.\fondo\  for $r\rightarrow
\infty $). But, quantum mechanically, $c$ is no more unit  in this
problem. We find that $c$ is related to the parameter $k$ of the
WZW model :
$$
c=\sqrt{{k\o{k-2}}}
\eqn\cluz
$$
and we reproduce the results of ref.[\DVV ]. In ref.[\DVV], the effect of
the renormalization of the speed of light is also present (although
it has not been noticed). Classically, before quantization, $D V V$
have $c=1$, but after quantization, since they choose $k = 9/4$,
they have $c=3$. This can be seen from the asymptotic behaviour of
the metric for $r\rightarrow \infty$. Classically, $dS^2
(r\rightarrow \infty) = dr^2 + d \Theta ^2$, and after CFT
quantization of the $SL(2,R)/U(1)$ model :
$$
dS^2_{DVV}\buildrel{r\to \infty}\over = 2(k-2)\left[
dr^2 + {k\o{k-2}}d\t^2 \right] = {4\o{c^2-1}}[dr^2 + c^2 d\t^2]
$$
Notice that in the classical limit $k\rightarrow \infty $ , $c$ takes
its classical value. The $S$-matrix of ref.[\DVV] is a particular case of
our results for $c=3$ :
$$
 S(p,c = 3) =S_{DVV}(p) = 2^{-2ip}{{\Gamma(ip)}\o{\Gamma(-ip)}}
 \left[{{\Gamma\left({1\o2}-2ip\right)}
\o {\Gamma\left({1\o2}-ip\right)}} \right]^2
\eqn\sdvv
$$
[The factor $ 2^{-2ip} $ is missing in ref. [\DVV ] ].
Notice that nothing special happens, however, at the conformal
invariant point $k=9/4$. The physical relevance of the  $k=9/4$
point in the  two dimensional black hole scattering matrix
is not clear. Also notice  that the presence or absence of conformal
anomalies is totally irrelevant for the black-hole  singularity
$ uv = 1 $ . [The conformal anomaly vanishes for $k = {9 \o 4}$ ].

The computation of the Hawking radiation follows directly from the
by now well known treatment of QFT in curved space-time and does
not present any particular feature here. The vacuum spectrum is a Planckian
distribution at the temperature $1/2\pi$.

We also quantize the system in the euclidean regime and exactly
compute the partition function of the  two dimensional stringy black
hole at temperature $\b^{-1}$. It is given by
$$
Z(\beta) = {1 \o {\pi}} \sum_{m,n=-\infty}^{+\infty}
\sqrt{ m^2 + (kn)^2 + 1} ~K_1\!\left({{\beta}\o{2k}}
\sqrt{ m^2 + (kn)^2 + 1}\right)
\eqn\zetai
$$
where $K_1(z)$ is a modified Bessel fuction. $Z(\b)$ is analytic for
all positive temperatures and hence no critical Hagedorn temperature
appears here.

In conclusion, the quantization of the center of mass of the string,
i.e. of the classical solution, with the requirement that $m^2=0$ ,
yields all the physics of the problem.We have obtained the full
exact quantum result without introducing any correction $1/k$ to the
space time metric, neither to the dilaton.

The  two dimensional CFT constructions can be avoided in a problem like the
two dimensional stringy black hole. They introduce a lot of technicality in
a problem in which all the physics can be described by the
straightforward quantization of a two-dimensional massless scalar
particle. Perhaps the CFT tools would be really necessary for the
problem of interacting (higher genus) strings in the  two dimensional black
hole background, problem, which unfortunately, has not been treated
until now.

Finally, let us comment about the pathological feature of the
renormalization of the speed of light. Normally, $c$ is never
affected by quantum corrections. Once one chooses units such that
$c=1$ classically, this value remains true in the quantum theory
(relativistic quantum mechanics, QFT, string theory, etc). In the
two dimensional stringy black-hole, it turns out that  $c = \sqrt{{k\o{k-2}}}$
 for the quantum massless particle described
by the string, whereas the wave equation in the metric \krus\
gives unit speed of propagation. Only for $k = \infty $ both speeds
coincide (but the conformal anomaly is present).

We think that this pathology is specific of the two
dimensional string dynamics in such curved background.
More precisely, this effect is to be traced back to
the dilaton background which couples with the
string only at the quantum level and as a surface effect (infinite
distance).

 \chapter{{\bf Strings in two dimensional geometries and dilaton backgrounds}}

The action for a string in a two dimensional space-time can be written as
$$\eqalign{
    {\cal S}  =& -{{1}\o{2 \pi }} \int d\s d\tau \sqrt{h} \left[ { {
h_{\mu\nu}(\s,\tau)} \o 2} \,
 G_{AB}(X) \,
 \partial^{\mu} X^A(\s,\tau) \, \partial^{\nu} X^B(\s,\tau)
- {{R^{(2)}}\o 4} \Phi (X) \right] \cr +& {\rm ~surface~ terms} ~.\qquad
\qquad \qquad \qquad A,B=1,2~;~\mu , \nu = 1, 2.\cr}
 \eqn\accion
 $$
where $h_{\mu\nu}$ is the metric on the world sheet, $ G_{AB}(X)$
 is the space time
metric, $R^{(2)}$ is the world sheet curvature and $\Phi (X)$ is the
dilaton field. $\Phi (X)$ and $G_{AB}(X)$ are constrained by the vanishing
of the beta functions. To one loop  level this implies
$$
R_{AB} = D_A D_B \Phi
$$
Classically, one has the string propagating in the background $G_{AB}$,
and the dilaton is decoupled from them.

We can always choose the conformal gauge on the world sheet, such that
$$
h_{\mu\nu}(\s,\tau)  =  \exp[\, \varphi(\s,\tau) \, ]\, \, {\rm diag}( -1, +1)
$$
Then,
$$
 \sqrt{h}\,  R^{(2)} = (\partial^2_{\tau}-\partial^2_{\s}) \varphi(\s,\tau)
\eqn\erre
$$
and
$$
    {\cal S}  = -{{1}\o{2 \pi }} \int d\s d\tau \left[ {1\o 2} \,
 G_{AB}(X) \,
 \partial_{\mu}X^A(\s,\tau) \, \partial^{\mu}X^B(\s,\tau)
- {1 \o 4}\, \partial^{\mu} \Phi (X)  \partial_{\mu}\varphi \right] .
 \eqn\accconf
 $$
Here $ \partial^{\mu}\a  \partial_{\mu}\beta = -{\dot \a}{\dot \beta}+
\a'\beta' $
 and we made a precise choice of the surface terms. Recall that
under a  Weyl transformation
$$
  h_{\mu\nu}(\s,\tau) \to e^{\l(\s,\tau)} \,   h_{\mu\nu}(\s,\tau)
 \eqn\weyl
 $$
the first term in eq.\accion\  is (classically) invariant. This is not the case
of the dilaton term that transforms as
$$
 \sqrt{h}\,  R^{(2)} \to  \sqrt{h}\,  R^{(2)} + \partial^2 \l
$$
A way out to have the Weyl invariance arises when the dilaton term is a
total divergence, and that needs $\partial_{\mu} \Phi = $ constant.

More precisely, for a genus zero euclidean world-sheet (dominant quantum
level) we can start from a flat metric on the complex plane
$$
dS^2 = ~ dz ~d{\bar z}
\eqn\metrhu
$$
and, map it into the euclidean world sheet $(\sigma, \tau^\prime )$ as
$$
z = e^{\tau' + i \s } ~,~~ 0 \leq \s < 2\pi~,~-\infty <
\tau^\prime < +\infty
\eqn\trafo
$$
The Lorentzian world sheet follows by Wick rotation, $\tau^{\prime} =
i\tau$ where $\tau$ is the (real) world sheet time. It follows from
eqs.\metrhu\ and \trafo\  that
$$
dS^2 = e^{2\tau'} \left[ d\tau'^2 + d\s^2 \right] =
e^{2i\tau} \left[ -d\tau^2 + d\s^2 \right]
\eqn\metbis
$$
That is,
$$
\varphi(\s,\tau) = 2 i \tau
\eqn\filou
$$
Therefore,
$$
 {1 \o 4}\, \partial^{\mu} \Phi (X)  \partial_{\mu}\varphi =
 -{i\o 2}{ \partial \Phi (X) \o {\partial\tau}}
$$
and the last term in eq.\accconf\  becomes a total derivative. It is
therefore irrelevant for the classical solutions. Thanks to the Weyl
invariance, the action eq.\accconf\  is $\varphi$ -independent and hence, real,
except for the surface term.

As is known, since the space time is two dimensional, we can always find
conformal coordinates $(U,V)$ where
$$
 G_{AB}(X) = e^{2\omega(U,V)}\pmatrix{ 0 &1 \cr 1 &0 \cr} ~ .
 \eqn\matriz
 $$
Then, the action \accconf\  takes the form
$$
    {\cal S}  = -{{1}\o{2 \pi }} \int d\s d\tau \left[  \,
   e^{2\omega(U,V)}
 \partial_{\mu}U(\s,\tau) \, \partial^{\mu}V(\s,\tau)
+  {i \o 2}\, {\partial \Phi (X) \o {\partial \tau} }\right] .
 \eqn\accuv
 $$
The equations of motion result given by
$$
\partial_+ \partial_- V + 2\, {{\partial \omega}\o {\partial V}}\,
\partial_+ V \, \partial_- V = 0 ~, \eqn\ecmovv
$$
$$
\partial_+ \partial_- U + 2 \, {{\partial \omega}\o {\partial U}}\,
\partial_+ U \, \partial_- U = 0 ~,
\eqn\ecmovu
$$
and the constraints are given by
$$
T_{\pm\pm} =  e^{2\omega(U,V)}\, \partial_{\pm}U \, \partial_{\pm}V = 0
\eqn\vincu
$$
where
$$
 \partial_{\pm} = {1 \o 2}( \partial_{\s} \pm  \partial_{\tau} )~~,~~
x_{\pm} = \s \pm \tau ~.
$$
Eqs.\ecmovv\ - \vincu\  posses the following solutions
\item{(i)} $$U = f(\s-\tau) \quad , \quad V = g(\s+\tau)
\eqn\soluno
$$
\item{(ii)}
$$U = f(\s+\tau) \quad , \quad V = g(\s-\tau)
\eqn\soldos
$$
where $f$ and $g$ are arbitrary functions of the indicated variables.
\item{(iii)}
$$ U = u_0 = {\rm constant ~ and}~ V = V(\s,\tau) {\rm
{}~is~ a~ solution~ of~eq. \ecmovv } \nxl
\eqn\soltre
$$
\item{(iv)}
$$ V = v_0 = {\rm constant ~ and}~ U = U(\s,\tau) {\rm
{}~is~ a~ solution~ of~eq. \ecmovu}
\eqn\solcua
$$
In the case (iii), in order to solve eq.\ecmovv\ with $U = v_0$, we set
$W = F(V)$ . Then,
$$
\partial_+ \partial_- W = F'(V) \left[\partial_+ \partial_- V +
{{F''(V) } \o { F'(V)}}\, \partial_+ V \, \partial_- V \right]
\eqn\doblw
$$
By choosing
$$
{{F''(V) } \o { F'(V)}} = 2 \, { {\partial \omega (u_0,V)}\o{ \partial V}}
\eqn\deduc
$$
eq.\ecmovv\ is fulfilled with $U = u_0$. Now, by integrating
eq.\deduc\  we find
$$\eqalign{
F(V) =& ~A \int_0^V dv   e^{2\omega(u_0 ,v)} + B~,  \cr
W =&  ~f(\s-\tau) +  g(\s+\tau)~,~ {\rm and~}~ V = F^{-1}(W) \cr}
\eqn\efe
$$
where $A$ and $B$ are arbitrary constants.

Similarly, for the solutions of type (iv) eq.\solcua\  we find
$$\eqalign{
 U =& ~G^{-1}(W)~{\rm where}~\partial_+ \partial_- W = 0~{\rm and} \cr
G(U) =& ~C \int_0^U du   e^{2\omega(u ,v_0)} + D~,  \cr}
\eqn\gee
$$
$C$ and $D$ being arbitrary constants.

Let us analyze the physical content of the solutions. The solutions of
type  (iii) and (iv) describe geodesics propagating on the
characteristies $U = u_0$ and $V = v_0$ respectively. We can make on these
solutions a conformal transformation  $ \s \pm \tau \to \phi_{\pm}(\s\pm\tau)$
making $W = \tau$.
Then, we obtain $\sigma$-independent solutions
$$
U = u_0 ~~~,~~~ V = F^{-1}(\tau)
\eqn\ucero
$$
Or,
$$
U = G^{-1}(\tau) ~~~,~~~V = v_0
\eqn\vcero
$$
where $F(V)$ and $G(U)$ are defined by eqs. \efe\ - \gee\ .

In other words, the solutions of type (iii) and (iv) reduce in an
appropriate gauge to the {\bf geodesics} for massless particles.

Let us discuss now the type (i) - solutions. In general, we can write
$$\eqalign{
U(\s-\tau) =& ~q_U + (\s-\tau)\, p_U + \sum_{n \neq 0}\a_{n}
\, e^{in(\s - \tau)} \cr
V(\s+\tau) =& ~q_V + (\s+\tau) \, p_V + \sum_{n \neq 0}\tilde\a_{n}
\, e^{-in(\s + \tau)} \cr}
\eqn\tipi
$$
Therefore,
$$
U(\s + 2\pi -\tau) - U(\s-\tau) = 2\pi \, p_U ~~,~~
V(\s + 2\pi +\tau) - U(\s+\tau) = 2\pi \, p_V
$$
If the physical space-time {\it does not} enjoy  periodic
directions, we must set $p_U = p_V = 0$. Then, we can locally gauge away
completely the $\alpha_n$ and $\tilde\a_{n}$-dependence  leaving no oscillator
degrees of freedom. This is exactely like a string in two
dimensional Minkowski space-time.

When {\it periodic directions do exist}, the situation is different. Let
us call $a$ and $b$ the periods in the coordinates $X = U - V$ and
$T = U + V$ respectively. Then, periodicity requires
$$
p_U = {{M b + N a} \o {4 \pi}} ~~,~~p_V = {{M b - N a} \o {4 \pi}}
\eqn\perio
$$
where $M$ and $N$ are integers. We can call them winding numbers. If
there is periodicity only on $X$, we should set $M b = 0$ (and
viceversa). That is, we can locally write the solutions as
$$
T(\s,\tau) = q_X + {{M b}\o{2\pi}}\s -{{N a}\o{2\pi}}\tau ~~,~~
X(\s,\tau) = q_T + {{N a}\o{2\pi}}\s -{{M b}\o{2\pi}}\tau
\eqn\xtpe
$$
In conclusion, in the absence of periodical directions we are left with
solutions of type (iii) and (iv). The $\sigma$-components can be
completely gauged away and we are left merely with (massless) point
particle geodesics. The only physical degree of freedom is the one
associated to the center of mass. That is, in the absence of transverse
dimensions in two dimensions, the strings has no excitations and hence,
it only describes one massless scalar particle. This is more a point
particle field theory than a string theory.

\chapter{{\bf String propagation in two dimensional black hole geometries}}

Let us now discuss the string propagation in the two-dimensional
black hole background of refs.[\bars -\Ed ]. In this case
$$
dS^2=k\left[ dr^2 - \tanh^2r \, dt^2 \right] ~,
{}~ 0<r<\infty ,~ -\infty < t < +\infty.
\eqn\metri
$$
where $k$ is a positive integer and the dilaton field takes the form
$$
\phi = \log (\cosh^2r)
\eqn\dila
$$
$k$ must be an integer in order to have $e^{i\cal S}$ univalued since it is
introduced as the coefficient of a multivalued WZW term [\Ed ].

Here, $(r,t)$ are Schwarzschild type coordinates related to Kruskal
coordinates $(u,v)$ by
$$
u = \sinh r ~ e^t~,~v=-\sinh r ~ e^{-t}~,~~ 0<r<\infty ,~ -\infty < t <
+\infty.
\eqn\krusk
$$
In terms of $(u,v)$, the length element eq.\metri\ takes the form
$$
dS^2=k~{{du dv} \o {1 - uv}}~,~~0<u,v< +\infty.
\eqn\kruska
$$
The hyperbola $u v = 1$ is a singularity of the space-time. (The
curvature is $R = {4 \o {\cosh^2 r}} = {4 \o {1-uv}})$.
The null lines $u v = 0$ (at which $r = 0$ and $t = \pm  \infty $)
correspond to the event horizon.

It is convenient to define the variable $r^{\star}$ by
$$
\sinh r = e^{r^\star} \qquad  -\infty < r^{\star} < +\infty.
\eqn\restr
$$
in terms of which the metric is written as
$$
dS^2 = k\; {{e^{2 r^{\star}}} \o {1 + e^{2 r^{\star}}}}~[(dr^{\star})^2
-dt^2]
\eqn\mret
$$
and we have simply
$$
u = e^{ r^{\star} + t} ~~,~~ v = -  e^{ r^{\star} - t}~.
\eqn\restb
$$
In the black hole background with {\it Lorentzian} signature, there are no
periodic dimensions. Then, the solutions of the string equations of
motion in the {\it Lorentzian} black hole background reduce to the type
(iii) and type (iv) solutions discussed in the previous section. That
is, they are just the geodesics of {\bf massless point} particles. Namely,
$$
u = u_0 = {\rm const.} \quad v = {1 \o {u_0}}
\left[ 1 - e^{u_0 (a\tau + b)} \right]
\eqn\geodeu
$$
Or,
$$
v = v_0 = {\rm const.} \quad u = {1 \o {v_0}}
\left[ 1 - e^{v_0 (a\tau + b)} \right]
\eqn\geodev
$$
That is,
$$
 r^{\star} = {1 \o 2} \log\left[ e^{u_0 (a\tau + b)} - 1 \right]~~,~~
 t = - {1 \o 2} \log\left[ {{e^{u_0 (a\tau + b)} - 1}\o {u_0^2}}
\right]
\eqn\geort
$$
It is interesting to discuss the relation between the world sheet
time $(\tau)$, and the physical-Schwarzschild $(t)$ or Kruskal $(T =
u + v)$ time. (Let us take the $u =$ const. case). We see that :
$$
v(\tau \to 0) \to A + B \tau \quad ,\quad t(\tau \to 0) \to -{1\o2}\log\tau
\eqn\tauc
$$
and
$$
v(\tau \to \infty) \to e^{u_0 a\tau} \quad ,\quad t(\tau \to \infty) \to
 -{1\o2} (a\tau + b)
\eqn\tauin
$$
We see that the world sheet time $\tau$ {\bf interpolates} between the
Kruskal time $(\tau \rightarrow 0)$ and the Schwarzschild time $(\tau
\rightarrow \infty)$. This is like strings in cosmological
backgrounds in which the world sheet time interpolates between
conformal time $(\tau \rightarrow 0)$ and cosmic time $(\tau
\rightarrow \infty )$. A similar relation holds also in gravitational
wave backgrounds [\ondpl ]. The logarithmic relation between Schwarzschild
time and $\tau$ for $\tau \rightarrow 0$ in eq.\tauc\  is exactly like
in de Sitter space between cosmic time and $\tau$. More generally,
$\tau \rightarrow \infty$ characterizes the time with respect to
which strings in $D$-dimensional curved spacetime oscillate,
then ingoing and outgoing
scattering states can be defined and the usual particle
interpretation can be applied. On the other hand, the $\tau
\rightarrow 0$ regime characterizes the time gouverning the string
evolution in the strong gravitational field regime where the string
does not oscillate in time.

It must be noticed that inside this two dimensional black hole, the
string takes an infinite proper time $\tau$ to approach the hyperbola
$u v = 1$ and then, never crosses it. In other words, the string
never reaches the space-time singularity. This is to be contrasted to
the higher dimensional case, in which for all $D > 2$, the string
reaches the singularity in a finite proper time $\tau$. Depending on
the singularity strength, the string crosses smoothly the space-time
singularity, or, alternatively, it is trapped by it [\ondpl ]. In the
$D$-dimensional black hole the string falls down into the space-time
$u v = 1$ singularity (in a {\it finite} time $\tau$), and classically, as
well as quantum mechanically , the string is trapped by the
singularity [\agune ] .

Another generic feature of string propagating in $D$-dimensional
curved backgrounds which {\it is lost} in the $D = 2$ case with Lorentzian
signature is the {\it string stretching}. As it is known
[\ondpl -\frolov ,\agune ], when strings
propagate in inflationary cosmological backgrounds and near
space-time singularities, the string proper length grows
indefinitely. This is accompagnied by a non oscillatory behaviour in
time, typical of this string instability. This is not present in
the $D = 2$ case with Lorentzian signature, since the transverse
oscillators are absent, and the $\sigma$-dependence is gauged away,
the string in $D = 2$ behaving like a point particle rather than an
extended object.

\chapter{{\bf Two dimensional Euclidean black hole. Cigar solutions and
Trumpet solutions}}

In the $D = 2$ black hole space-time with Lorentzian signature, in
which compact directions are absent, the only string physical degree
of freedom is the center of mass (point particle). In the Euclidean
signature case, however, compact dimensions are present and the
$\sigma$-dependence remains. Let us consider now the solutions in the
euclidean regime. The Schwarzschild time is purely imaginary, namely
$$
t = i \Th \quad{\rm i. e.}\quad
dS^2=k\left[ dr^2 + \tanh^2r \, d\Th^2 \right] ~,
\eqn\tima
$$
and the Kruskal coordinates are
$$
u = \sinh r ~ e^{i\Th}~,~v=-\sinh r ~ e^{-i\Th}~,
{}~ 0<r<\infty ,~ 0 < \Th < 2 \pi.
\eqn\kruske
$$
The "euclidean time" $\Theta$ is periodic with period $2\pi$.

For solutions of the type (i) or (ii) [eqs. \soluno\ or \soldos ] in the
euclidean regime, periodicity in $\sigma$ requires
$$
\s = i {\hat \s} = i {{\Th} \o n}
$$
that is, the world sheet signature must also be euclidean  with  $ {\hat \s}$
playing the r\^ole of the euclidean time - (i.e., $t = n \s$).

The solution  is given by
$$
u = e^{n(\tau + i {\hat \s})}\quad,\quad v = - e^{n(\tau - i {\hat \s})}
\eqn\soleu
$$
that is,
$$
\sinh r = e^{n\tau}~~,~~i. e.~~  r^{\star} = n\tau~~;~~\Th = n {\hat \s}~~,
\eqn\soleud
$$
with the dilaton given by
$$
\Phi = \log[1 + e^{2n\tau}] ~.
\eqn\dileu
$$
This is a non singular real manifold with metric
$$
 dS^2 = k \;{{n^2}\o{1+e^{-2n\tau}}}~(d\tau^2+d{\hat \s}^2)
\eqn\cigarr
$$
For $\tau \rightarrow \infty $ the metric is flat.

$\Theta = n {\hat \sigma} $ are winding configurations. They correspond to
the so-called "momentum modes", in the terminology of the literature
(see ref.[\DVV]), since here, ${\hat \s}$ is an imaginary time. The physical
interpretation of these modes is a string winding $n$ times along the
angular $\Theta$ direction, lying in the semi-infinite cigar from  $r = 0$
to $r = \infty $. At $\tau = - \infty$, the string is at $r = 0$ (the
horizon is shrinked to a point in the euclidean space). Notice, that
this is a {\it rigid string}  configuration staying around the cigar
without {\it any oscillations}. In $D = 2$, $e^{in{\hat \s}}$
 does not describes  string
fluctuations as in the $D > 2$ cases, thus the name {\bf modes} ("winding
modes" or "momentum modes") referring to strings in $D = 2$ are
misleading.

Also notice that in the $D = 2$ euclidean regime, the string world
sheet {\it covers} completely the target space ; world sheet and physical
space became the same ; (for $n = 1$, both are identified). This is
characteristic of strings in $D = 2$ compact spaces. Recall that in
$D = 2$ de Sitter space [\desi ] , the general solution of the string equations
of motion describes a string winding $n$ times around de Sitter
universe (the circle $S^1$) and evolving with it. In that case, the
winding string solution naturally exists in the {\it Lorentzian} signature
regime in both target and world sheet  space-times ; (it can be also
extended to the Euclidean regime). The winding string de Sitter
solution  is given by [\desi ]
$$
dS_{deSitter}^2 = {1\o{H^2}}\left[-du^2 + \cosh^2u~dv^2 \right]
\eqn\desim
$$
with $\quad v=n\s , \qquad u = \log(\tan{{n\tau}\o2})~~ ,  \qquad 0<\s\leq 2
\pi~,
{}~~~~ 0< \tau \leq {{2\pi}\o n}$.

That is,
$$
dS_{deSitter}^2 = {{n^2}\o{H^2\sin^2n\tau}}(d\s^2~-~d\tau^2)
\eqn\desih
$$
[here $(u, v)$ are the coordinates on the de Sitter hyperboloid].

In the black hole background, winding string solutions are only
possible in the Euclidean signature regime.

It must be noticed that the transformation
$$
r = s + {{i\pi}\o2} ~~,~~ \Th = \Th
\eqn\trafo
$$
maps the manifold eq. \tima\  into the dual or "trumpet" manifold
$$
dS^2 = k \left[ ds^2 + \coth^2s ~ d\Th^2 \right]
\eqn\tromp
$$
This is a real manifold with curvature singularity at $s = 0,~ (u v =
1)$.
The Kruskal coordinates are complex conjugate one to another, i.e.
$$
u = \cosh s ~ e^{i(\Th + {{\pi}\o 2})}~~~,~~~
v  = \cosh s ~ e^{-i(\Th + {{\pi}\o 2})}
\eqn\krusc
$$
The string solution eqs.\soleu\ - \soleud\  transforms under eq.\trafo\ into
the
different real solution
$$
\cosh s = e^{n\tau}~~~,~~~\Th = n {\hat \s}
\eqn\tromd
$$
(the dilaton here is $\Phi = \log[1-e^{2n\tau}]$).
That is,
$$
u = e^{n(\tau + i\hat \s) + i\pi/2}~~,~~
v =  e^{n(\tau - i\hat \s) - i\pi/2}
\eqn\trom
$$
For this solution, the metric is
$$
 dS^2 = k\; {{n^2}\o{1-e^{-2n\tau}}}~(d\tau^2+d{\hat \s}^2)
\eqn\tromp
$$
The proper string length (between two points $\hat \s$ and $\hat \s
+ d\hat \s $ at fixed $\tau$) is given by
$$
\Delta S_{trumpet} = \sqrt{k}\; {n\o{\sqrt{1-e^{2n\tau}}}} ~d{\hat \s}
\eqn\strom
$$
Near the curvature singularity $s\rightarrow 0 ~ (i.e.~ \tau \rightarrow
0),~ \Delta S$  blows up, the string stretches indefinitely,
corresponding to the location of the string in the open extremity of
the trumpet.

\section{Cosmological solutions}

The cosmological version of the black hole solution eq.\metri\  is just
obtained by setting
$$
r=iT \quad,\quad t = x \eqn\wick
$$
in the original metric. The cosmological background is described by
$$\eqalign{
dS^2&=k\left[ -dT^2 + \tan^2T \, dx^2 \right] ~, \cr \Phi &= -\log\sin T~,
{}~ 0<T<\pi/2 ,~ -\infty < x < +\infty.}
\eqn\cosm
$$
which is singular in $uv = 1$ , i. e.$~T = \pi/2$. Here
$$
u = i \sin T ~ e^x \quad , \quad v = -i \sin T ~ e^{-x}
\eqn\coorc
$$
It is conveniently written in the variable $T^{\star}$
$$
\sin T = e^{T^{\star}} \quad ,\quad  -\infty < T^{\star} < 0
$$
i.e.
$$
dS^2 = {1 \o {1 - e^{-2 T^{\star}}}} ~\left[ (- d T^{\star})^2 + dx^2 \right]
\eqn\meco
$$
The analysis of the string solutions done in section IV for the
black hole, holds here with the corresponding mapping eq.\wick. There
are no solutions describing a string winding $n$ times in the
Lorentzian regime. In the Euclidean regime
$$\eqalign{
x =& ~i \Th ~~,~~0 \leq \Th < 2 \pi,\cr
-dS^2 =& ~{1 \o {1 - e^{-2 T^{\star}}}} ~
\left[ ( d T^{\star})^2 + d\Th^2 \right]~. \cr}
\eqn\mecoe
$$
We have the solution
$$
 T^{\star} = n \tau ~~,~~ \Th = n {\hat \s}~~,~~-\infty < \tau \le 0
{}~,~~ 0 \le  {\hat \s} < {{2\pi}\o n}
$$
That is
$$\eqalign{
u ~~~=& ~~~ i ~ e^{T^{\star} + i \Th}~~~ =~~~ i ~ e^{n(\tau + i {\hat \s})},
\cr
v~~~ =& ~~ -i ~ e^{T^{\star} - i \Th}~~~ =~~ -i ~ e^{n(\tau - i {\hat \s})}, }
\eqn\coord
$$
describes a string winding $n$ times in the euclidean cosmology.
This solution yields
$$
 dS^2 = k \; {{n^2}\o{1-e^{-2n\tau}}}~(d\tau^2+d{\hat \s}^2)\; .
\eqn\costr
$$
The string length blows up at $\tau \rightarrow 0$.

The cosmological version of the trumpet manifold is obtained by
setting
$$\eqalign{
T =& ~ -i U + {\pi /2} ~~, ~~{\rm i. e. } \cr
 dS^2 =& ~ k\;[  ds^2 + \coth^2 U ~ d\Th^2 ] = k \;
{1 \o {1 - e^{-2 U^{\star}}}} ~
\left[ ( d U^{\star})^2 + d\Th^2 \right] \cr}
\eqn\cotro
$$
with
$$
u = \cosh U ~ e^{i \t} = e^{ U^{\star} + i\t}~~,~~
v = \cosh U ~ e^{-i \t} = e^{ U^{\star} - i\t}
$$
The winding string solution here is given by
$$
 U^{\star} = n \tau ~~,~~ \Th = n {\hat \s}~~,~~-\infty < \tau < 0
{}~,~~ 0 <  {\hat \s} < {{2\pi}\o n}
\eqn\enrrc
$$
The solutions found in sections (IV) and (V) exhaust {\bf all} the real string
solutions in the lorentzian and euclidean regimes. The
$\sigma$-dependent solutions exist only in the euclidean regime.
These solutions are of the type (i) and (ii) [eqs.\soluno\ and
\soldos ].
There are no real $\sigma$-dependent euclidean solutions of the type
(iii) and (iv) [eqs.\soltre\ and \solcua\ ].

There are two types of euclidean string solutions : cigar solutions and
trumpet solutions. These correspond to a string winding $n$ times in
the cigar and in the trumpet manifolds respectively. In the
cosmological version of the black hole manifold (in the euclidean
regime) the string solutions are the same as the trumpet solutions.

\chapter{{\bf String quantization in the two dimensional black-hole}}

In the Lorentzian two dimensional black hole background eqs.\metri\ - \dila\
the action eq.\accconf\  takes the form
$$
    {\cal S}  = -{{1}\o{2 \pi }} \int d\s d\tau \left\{ {k\o 2} \,
\left[{\dot r}^2 - r'^2 - \tanh^2r ~({\dot t}^2 - t'^2) \right]
- i~{\dot r} \tanh r \right\}
\eqn\accan
$$
where we also used eq.\metbis\  for the world sheet metric. Notice that
the dilaton field leaves only a surface term analogous to a
background charge in conformal field theory [\volodia ].

The Hamiltonian associated to the action \accan\   has the form
$$
{\cal H} = {1 \o {2k}}\left[ (p_r - i \tanh r )^2 -
p_t^2 \coth ^2r \right] +  {k\o 2} \left( r'^2 - \tanh^2r~ t'^2
\right)
\eqn\hamu
$$
As previously discussed, in absence of periodic directions in the
space-time, the parameter $\sigma$ can be classically gauged away.
Since this is the case for black holes with Lorentzian signature, we
can drop the second term in eq.\hamu\  and take
$$
{\cal H} = {1 \o {2k}}\left[ (p_r - i \tanh r )^2 -
p_t^2 \coth ^2r \right]
\eqn\hamd
$$
Let us now consider the quantum theory. We will follow the spirit of
ref.[\nos ] where one starts from the center of mass motion and then,
takes into account the $\sigma$-dependence as fluctuations around.
Here, in $D = 2$, there are no physical fluctuations, so we can
expect that the quantum version of the center of mass hamiltonian
eq.\hamd\  should give the full description of the model.

In order to quantize this system, we use the canonical prescription
$$
p_r \to {\hat p_r} = - i {{\partial} \o {\partial r}}~~~,~~~
p_t \to {\hat p_t} = - i {{\partial} \o {\partial t}}
\eqn\cuan
$$
Now, eq.\hamd\ suffers from ordering ambiguities at the quantum level.
We choose the following ordering prescription. The classical piece
$ p_r^2 - p_t^2 \coth ^2r $ in eq. \hamd\  is replaced by
the D'Alambertian in this space-time :
$$
 p_r^2 - p_t^2 \coth ^2r \to  \partial^2 = {1 \o {\tanh r}}
 {{\partial} \o {\partial r}}(\tanh r  {{\partial} \o {\partial r}})
- \coth^2r ~{{\partial^2} \o {\partial t^2}}
\eqn\dala
$$
The operator $\partial^2$ is self adjoint with respect to the integration
measure $\sqrt{G} = k \tanh r$.

The double product $-2ip_r \tanh r$ in eq.\hamd\  is ordered symmetrically
$$
-2ip_r \tanh r \to -i( \tanh r ~{\hat p_r} +  {\hat p_r} \tanh r )
\eqn\orsi
$$
The last term $- \tanh^2 r$  does not present ordering problem. We find in this
way
$$
{\cal H} \to \hat{\cal H}= - {1 \o {2k}}\left(
{{\partial^2} \o {\partial r^2}}+ (\coth r + \tanh r )
{{\partial} \o {\partial r}}- \coth^2r ~{{\partial^2} \o {\partial t^2}}
\right)
\eqn\hami
$$
Let us write
$$
{\cal H} \to \hat{\cal H}= {1 \o {2k}} h \qquad,\qquad \partial_t = -iE
$$
then,
$$
h = -\left[{{d^2}\o{dr^2}} + 2 \coth(2r) {d\o{dr}} + E^2 \coth^2r\right]~~,
\eqn\hamt
$$
and the eigenvalue problem
$$
h \Psi = \l \Psi ~~~,
\eqn\auto
$$
yields
$$
\Psi'' +  2 \coth(2r) \Psi' + ( E^2 \coth^2r + \l) \Psi = 0
\eqn\ecau
$$
The constant $\lambda$ describes the zero point quantum fluctuations
and it will be determined below ; (the prime ' stands  for ${d\o
{dr}}$).
 Eq.\ecau\ describes the interaction of the tachyon with the black hole
geometry. It is convenient to set here
$$
\Psi = {{\chi} \o {\sqrt{\sinh2r}}}
\eqn\fuxi
$$
then, eq.\ecau\  reads
$$
\chi'' + \left[ {{E^2+ {1 \o 4}}\o {\sinh^2 r}} - {1 \o { 4 \cosh^2
r}}
+ E^2 +\l - 1 \right] \chi = 0
\eqn\schr
$$
which is a Schr\"odinger type equation with the energy-dependent
effective potential
$$\eqalign{
V_{eff} =& -{{E^2+ {1 \o 4}}\o {\sinh^2 r}} + {1 \o { 4 \cosh^2
r}} - E^2 - \l + 1 + E \cr
V_{eff} \buildrel{r\to0}\over =& -{{E^2+1/4} \o{r^2}} + O(1) ~,\cr}
\eqn\potef
$$
For $r \rightarrow 0$, this is a strongly attractive singular
potential $-{{\g}\o{r^2}}$ .
 Since $\g > {1\o4}$ there is absorption (fall into the event horizon
$r = 0$) for all energy E, even $E = 0$. This is exactly like the
horizon behaviour of the effective potential describing the
interaction of the black hole with a free scalar field in $D = 4$, where
$\g = E^2+1/4 $ too.

The Hamiltonian eq.\hamt\ , is the same Hamiltonian $L_0$ as in ref.[\DVV ]. In
order to compare with ref.[\DVV ], (hereafter referred as DVV), please
notice that
$$
r_{DVV} = 2 r ~~,~~E = 2 \omega_{DVV}
\eqn\rdvv
$$
Then, eq.(4.11) in DVV reads
$$
L_0^{DVV}=-{1\o{k-2}}\left[
{{\partial^2} \o {\partial r^2}}+ (\coth r + \tanh r )
{{\partial} \o {\partial r}}- \left(\coth^2r-{2\o k}\right)
 {{\partial^2} \o {\partial t^2}}\right]
\eqn\eleo
$$
and the condition $(L_0 - 1 ) \chi$  yields
$$
\left[{{d^2}\o{dr^2}} + {{E^2+ {1 \o 4}}\o {\sinh^2 r}} - {1 \o { 4 \cosh^2
r}} + E^2 + 1 + 4(k-2) \right] \chi = 0
\eqn\vincu
$$
$\hat{\cal H}$ in  eq. \hami\  and the $L_0^{DVV}$ only differ by
additive and multiplicative
constants. As operators they are identically the same.

Let us now determine the constant $\lambda$. It can be determined from
the behaviour at $r \rightarrow \infty$ of the solution.
For $r\rightarrow \infty$ , from
eq.\ecau\  we have
$$
\Psi'' +  2 \Psi' + ( E^2  + \l) \Psi = 0
\eqn\ecaud
$$
And then, (including the time dependence of the total solution) :
$$
\Psi(r\to \infty , t) = e^{i(pr-Et)}
\eqn\ondpl
$$
with  the momentum $p = \sqrt{E^2 + \l - 1}$.
The parameter $\lambda$ becomes determined by requiring the tachyon to be
massless. In order to describe a relativistic and massless particle,
(the string ground state particle is massless in $D = 2$ due to the
absence of transverse dimensions), it must be $E = c p$, that is
$\l - 1 = (1 - c^2) p^2 $.
For comparison with DVV, notice that for $r\rightarrow \infty$, DVV
have
$$
\Psi'' +  2 \Psi' +  E^2(1 - {2 \o k } ) \Psi = 0
\eqn\ecdvv
$$
which corresponds to have $\l = 1 - 2 E^2 /k$.
That is, eqs.\ecaud\ and \ecdvv\  yield the relation
$$
c=\sqrt{{k\o{k-2}}}
\eqn\vluz
$$
We will comment about the interpretation of $\lambda$ and this relation
below.

In order to describe the solutions of eq.\ecau , it is useful to define
the variable
$$
z \equiv \cosh^2r
$$
and to express $\Psi$ , as
$$
\Psi_E(z)= (z-1)^{iE/2}~\psi_E(z)
\eqn\fpsi
$$
Then, eq.\ecau\  reads
$$
z(1-z)\psi_E(z)''+ [ 1 - z(iE + 2) ]\psi_E(z)' - \left({{iE}\o2}+
{{\l}\o4} \right)\psi_E(z) = 0
\eqn\ecpsi
$$
whose solutions are the standard hypergeometric functions.

The general solution is given by
$$
\Psi(r,t) = \left[\; A \; U_{E,c}(r) + B \;  U_{E,c}(r)^{\ast} \;
\right]e^{-iEt}
\eqn\solps
$$
where $A$ and $B$ are arbitrary constants and
$$
 U_{E,c}(r)=(\sinh r)^{- iE}
F\left({1\o2}- {{iE} \o 2} (1+ {1\o c}),{1\o2}- {{iE} \o 2} (1- {1\o c});
1- iE \, ;-\sinh^2r \right)
\eqn\solhi
$$
These solutions describe the scattering of the (massless) tachyon by the black
hole geometry. The solutions $U_{E,c}$ and $U_{E,c}^{\ast }$ exhibit the
typical behaviour of the wave functions near the black hole horizon,
namely
$$
 U_{E,c}(r)\, e^{-iEt}\; \buildrel{r\to0}\over = e^{-iE(t + \log r)} \quad ,
\quad
 U_{E,c}(r)^{\ast}\, e^{-iEt} \; \buildrel{r\to0}\over = e^{-iE(t - \log r)}
\eqn\horiz
$$
$U_{E,c}(r,t)$ corresponds to have purely incoming particles at the
(future) horizon. $U_{E,c}^{\ast}$ describes purely outcoming particles from
the (past) horizon. The solution $U_{E,c}$ describes the physical
process in which the tachyon field coming in from spatial infinity is
partially reflected and partially absorbed by the black hole. The
asymptotic behaviour of $U_{E,c}(r,t)$ for $r\rightarrow \infty $ is
given by
$$
 U_{E,c}(r)\; e^{-iEt} \; \buildrel{r\to \infty }\over = e^{-iEt} ~
\left[ e^{-iEr/c } + S(E,c) e^{iEr/c} \right].
\eqn\infin
$$
{}From eq.\solhi\ we obtain for the scattering matrix $S(E,c)$ :
$$\eqalign{
 S(E,c) &= 2^{-2iE/c}{{\Gamma(iE/c)}\o{\Gamma(-iE/c)}}
 \left[{{\Gamma\left({1\o2}-{{iE}\o2}(1+{1 \o c})\right)}
\o {\Gamma\left({1\o2}-{{iE}\o2}(1-{1 \o c })\right)}} \right]^2 \cr}
\eqn\mates
$$
or, in terms of $p$  eqs.\maese\ .
Notice that the S-matrix factorizes into two parts, one which is a pure
phase, equal to the coulombian S-matrix, and another one which
describes both absorption and reflexion, namely
$$
 S(p,c) = S_{coul}(p)~{\tilde S(p,c)}
$$
where
$$\eqalign{
 S_{coul}(p) &= 2^{-2ip}{{\Gamma(ip)}\o{\Gamma(-ip)}} =
2^{-2ip} e^{2i{\rm arg}\Gamma(ip)} \cr
{\tilde S(p,c)} &= \left[{{\Gamma\left({1\o2}-{{ip}\o2}(c+1)\right)}
\o {\Gamma\left({1\o2}-{{ip}\o2}(c-1)\right)}} \right]^2 \cr}
\eqn\matrs
$$
$S_{coul}$ takes into account the large $r$ interaction while
  ${\tilde S}$ describes the genuine black hole features. Notice that
$$\eqalign{
| S_{coul}(p)| &= 1 \cr
| {\tilde S(p,c)}| &= \left[{{\cosh{{{\pi p}\o2}(c-1)}}\o
{\cosh{{{\pi p}\o2}(c+1)}}}\right]^2 ~<~1~,\cr}
\eqn\modul
$$
which describes the absorption by the black hole.
Also notice that $S(E,c)$ exhibits imaginary poles. An infinite
sequence of purely imaginary poles occur for the values $i p = n~
(n=0,1,...)$ which are like the Coulombian-type poles (describing the
relativistic hydrogen bound-state spectrum). In addition, the part
$ {\tilde S(E,c)}$ exhibits an infinity sequence of purely imaginary poles at
$i p (c+1) = 2n+1, n=0,1,2...$.

It must be stressed that the scattering matrix $S(E,c)$ depends on two
parameters $E$ and $c$ . That is, for each energy we have a
monoparametric family of well defined solutions depending on the
parameter $c$. For each value of $c$, we have a different S-matrix.
It must be noticed that $c$ is a purely quantum mechanical parameter
which is not fixed at this level by any special requirement, $c$ is a
physical parameter which accounts for a renormalization of the speed of
light. At the classical level, $c$ can take any value, (we have taken
$c = 1$), the physics does not depend on it. But, quantum mechanically,
$c$ is not more equal to 1 in this problem.

The parameter $c$ is related to the parameter $k$
of the conformal field theory construction,  $ c=\sqrt{{k\o{k-2}}}$
 [see eq.\vluz ]. In
particular, for $k = 9/4, ~i. e. ~ c = 3$ we reproduce the results
of ref.[\DVV] . In this respect,
notice that in ref.[\DVV] , the effect of the renormalization of the speed
of light is also present, (although it has not been noticed there).
Classically, before quantization, DVV have $c=1$, and
after quantization (since they choose $k=9/4$), they have $c=3$. This
can be seen from the asymptotic behaviour of the metric for
$r\rightarrow \infty $.

Classically,
$$
dS^2=k\left[ \; dr^2 + \tanh^2r \, d\Th^2 \; \right] ~
\buildrel{r\to \infty }\over = \; dr^2 + d\Th^2 ~
$$
After quantization of the euclidean black hole CFT  coset
$SL(2,R)/U(1)$ model :
$$\eqalign{
dS^2_{DVV}=& ~{1 \o 2} (k-2) \left[ ~dr_{DVV}^2 \; + \; \beta^2(r_{DVV}) ~
d\Th^2 ~ \right]~,\cr
 {\rm where ~~~~}
\beta(r) =& ~ 2 \left(\coth^2{r \o 2} - {2 \o k} \right)^{-1/2} ~~~{\rm i.
e.}~~,~~ \beta(\infty) = {2 \o {\sqrt{1 - 2/k}}}~, \cr
dS^2_{DVV} \buildrel{r\to \infty }\over =
&  ~{1 \o 2} (k-2) \left[~dr_{DVV}^2 \; +
\; {{4k}\o {k-2}} \;  d\Th^2 ~ \right]~,\cr}
\eqn\metdvv
$$
Recall that $r_{DVV} = 2 r$ , i. e.
$$
dS^2_{DVV} \buildrel{r\to \infty }\over =   ~2 (k-2) \left[\; dr^2 \;+\;
 { k \o {k-2}}\;   d\Th^2 \; \right] =
{4\o {c^2 -1}} \left[ \; dr^2 \; + \; c^2 \,  d\Th^2 \; \right]
\eqn\mdvva
$$
We see that in ref. [\DVV] , it also appears that $c^2 = { k \o {k-2}}$
  in particular there, $k_{DVV} = 9/4 $.
Notice that nothing special happens at the conformal invariant point $k
= 9/4$. The physical relevance of the $k = 9/4$ point in the two-dimensional
black hole scattering matrix is not clear. The S-matrix of ref.[\DVV ]  is a
particular case of our results for $c=3$. The $S$-matrix, eq.(5.22) of
ref.[\DVV ] is given by
$$
S_{DVV}(p) = 2^{-2ip}{{\Gamma(ip)}\o{\Gamma(-ip)}}
 \left[{{\Gamma\left({1\o2}-2ip\right)}
\o {\Gamma\left({1\o2}-ip\right)}} \right]^2
\eqn\sdvvb
$$
[The factor $ 2^{-2ip} $ is missing in ref. [\DVV ] ].
{}From eq.\matrs\ , we see that
$$
S(p, c=3) =S_{DVV}(p)
\eqn\final
$$
In conclusion, in this problem, the quantization of the {\it center of mass
of the string}, that is the quantization of the classical solution,
together with the requirement that $m^2 = 0$, yields {\it all} the physics
of the problem. [In this sense, this is like the light cone gauge for
strings in flat spacetime for D=26 ]. We have obtained the complete
quantum result without introducing any correction $1/k$ to the
space-time metric, neither to the dilaton field.

The $2$-dimensional conformal field theory construction can be
avoided for such a simple problem like the  two-dimensional stringy black
hole. They introduce a lot of technicality in a problem in which all
the physics is described by the standard quantization of a $2$
dimensional massless scalar particle. Perhaps, the CFT tools
would be really necessary for the problem of interacting (higher
genus) strings in the  two dimensional black hole background, a problem
which, unfortunately, has not been treated until now.

Finally, let us comment about the only non trivial physical feature,
perhaps, appearing in this two dimensional stringy black hole
problem, namely the renormalization of the speed of light. Normally,
the speed of light is never affected by quantum corrections. That is,
once one chooses units such that $c = 1$ classically, this value remains
true in the quantum theory (relativistic quantum mechanics, or
quantum field theory or string theory, etc...). In the present case,
$c = 1$, for the classical propagation in the background eq.\fondo ),
while  $c=\sqrt{{k\o{k-2}}}$ for the quantum propagation of the massless
particle
described by the string. We think that this pathology is specific of
the two dimensional string dynamics in such curved backgrounds.
Only for $k \to \infty$, the pathology disappears (but the conformal
anomaly is present). In that case:
$$\eqalign{
S(p,c=1) =& ~S_{coul}(p)\; {1\o{\pi}}
 \; \left[\Gamma\left({1\o2}-ip\right) \right]^2 \cr
|S(p,c=1)| =& ~{1 \o { \cosh^2 {\pi p}}} \cr}~.
\eqn\ceuno
$$

\section{Quantum Partition Function}

Let us now consider the quantum statistical mechanics for this system
by computing its partition function  $Z = Tr \left[ e^{-\b \hat{\cal H}_E}
\right] $ , where $ \hat{\cal H}_E $ is the euclidean version
of the Lagrangian in eq.\accan\ . In the euclidean regime ( $t = i \Th
$) the variable $\Th$ is periodic with period $2\pi$ and in
an appropiate conformal gauge, generic
configurations take the form
$$
\Th (\s ,\tau) = n {\hat \s} + {\hat \Th}(\tau)
\eqn\gtet
$$
The whole $\s$-dependence can be gauged
away from $r$ . Thus, using eq.\gtet\ , we find an extra term in eq.\accan\ :
$$
{ 1 \o 2} k n^2 \tanh^2r
$$
This term plus the Wick rotated version of $ \hat{\cal H}$ [ eq.\hami\ ]
yields the euclidean hamiltonian
$$
 \hat{\cal H}_E = {1 \o {2k}}\; h_E
$$
with
$$
h_E = -{{\partial^2}\o{\partial r^2}} - 2 \coth(2r)\,
{\partial\o{\partial r}} - \coth^2r\;{{\partial^2}\o{\partial {\Th}^2}}
+(kn)^2 \tanh^2 r
\eqn\hame
$$
The eigenfunctions of $h_E$ take the form
$$
W(r, \Th) = e^{im\Th} ~ w_{m,n}(r)
$$
where $ m \epsilon {\cal Z}$ in order to have $2\pi$ periodicity and
$ w_{m,n}(r)$ is a solution of the eigenvalue equation
$$
\left[ -{{d^2}\o{dr^2}} - 2 \coth(2r)
{d \o{dr}} + m^2 \coth^2r
+(kn)^2 \tanh^2 r \right] w_{m,n}(r)~ = ~\Lambda ~ w_{m,n}(r)
\eqn\eigeu
$$
Regular solutions behave as
$$
 w_{m,n}(r) \buildrel{r\to 0}\over = r^{|m|}
\quad , \quad   w_{m,n}(r) \buildrel{r\to \infty}\over =
e^{-r} \left[ e^{ipr} + S_{m,n}(p) \right]
\eqn\asyn
$$
where $p~  \epsilon~ {\cal R}$ is the euclidean momentum. The eigenvalues
$\Lambda$ of $h_E$ on such eigenfunctions are
$$
\Lambda_{p,m,n} = \sqrt{ p^2 + m^2 + (kn)^2 + 1} ~~~~~n,m ~ \epsilon~
{\cal Z}~,~ p  ~ \epsilon ~{\cal R}.
$$
Then, the partition function results:
$$
Z(\beta) = Tr\left[ e^{-\b \hat{\cal H}_E} \right]
= \sum_{m,n=-\infty}^{+\infty}
$$
The integral over $p$ can be computed with the result
$$
Z(\beta) = {1 \o {\pi}} \sum_{m,n=-\infty}^{+\infty}
\sqrt{ m^2 + (kn)^2 + 1} ~K_1\!\left({{\beta}\o{2k}}\sqrt{ m^2 + (kn)^2 +
1}\right)
\eqn\zeta
$$
There is a manifest duality invariance $m \leftrightarrow kn$ .

For large $\b$ (low temperatures), $Z(\b)$ decreases exponentially as
$$
Z(\b) \buildrel{\b \to \infty }\over = \sqrt{{k \o {\pi \b}}}~
e^{-{{\beta}\o{2k}}}
$$
due to the presence of  a gap in the eigenvalues :$\Lambda_{p,m,n} \ge 1$.
For high temperatures, $Z(\b)$ grows as $\b^{-3}$ :
$$
Z(\b) \buildrel{\b \to 0 }\over = {{16 k^2} \o {\b^3}}
$$
Notice that  $Z(\b)$ is analytic for all $\infty \ge \b > 0$
and hence no phase transition shows up here for finite temperatures.
The absence of a Hagedorn critical  temperature
is due to the lack of transverse string modes in two dimensions.

\chapter{{\bf Hawking radiation}}

The computation of the Hawking radiation follows directly from the
standard treatment of the by now well known QFT in curved space
time and does not present any particular problem here. For the sake
of completeness we point out the essential points to be considered
and which allow to avoid unnecessary
complications. The formulation of quantum (point particle) field
theory, as well as quantum string theory in non trivial spacetime
exhibits new fundamental features with respect to the usual
formulation in flat space time: (i) the possibility for a given field
or string theory to have different alternative well defined ground
states, and thus different Fock spaces or "sectors" of the theory.
(ii) the presence of intrinsic statistical features (temperature,
entropy) arising from the non trivial structure (geometry, topology)
of the space time and not from a superimposed statistical description
of the quantum matter fields or strings.

An important step in field as well as string quantization is the
definition of positive frequency states and its associated ground
state. Different possible choices of the physical time lead to
different alternative positive frequency basis.

The modes $U_{E,c}(r,t)$ (see eq.\solhi ) define positive frequencies with
respect to the Schwarschild time and are the analogous of the Rindler
modes for an uniformely accelerated observer $r =$ const. Creation and
anhihilation operators ${a_E ,~ a_E^{\ast}}$ are associated to these modes
which define the ground state $|0_S>$ or "Schwarzschild'' ground state.

The modes $U_{E,c}$ are however, a linear combination of positive
and negative frequency modes with respect to the Kruskal time $T$.
For a locally inertial observer near the horizon or Kruskal
observer, the positive frequency basis is
$$
\varphi_k \buildrel{r\to 0}\over = \{ {1 \o {2\sqrt{\pi k}}} ~ e^{-iku}~,
 {1 \o {2\sqrt{\pi k}}} ~ e^{-ikv} \}~,~~k~>~0~,
$$
with respect to which the global or Kruskal  ground state is defined
$\a_k \, |0_K> = 0$ .
The Bogoliubov coefficients relating such two basis, namely,
$$
\varphi_k = {\cal A}_{kE}~U_{E,c}(r,t) +  {\cal B}_{kE}~U_{E,c}^{\ast}(r,t)
\quad [~{\rm or}~ \a_k =  {\cal A}_{kE}~~a_E  +  {\cal B}_{kE} ~~a_E^{\ast}~]~,
$$
are defined by the scalar products
$$
 {\cal A}_{kE} ~ = ~ <U_E , \varphi_k^{\ast}>~~,
{}~ {\cal B}_{kE} ~ = ~ <U_E ,\varphi_k >~,
$$
and explicitly given by
$$
 {\cal A}_{kE} = e^{\pi E} ~ {\cal B}_{kE} \quad , \quad
 {\cal B}_{kE} \buildrel{k \to \infty }\over =
 - {{\sqrt{E} ~ k^{iE}} \o {2\pi\sqrt{k}}} ~ \Gamma(-iE) \;
e^{-\pi E/2}
$$
The behaviour of the basis $\{ U_E \}$ and $\{ \varphi_k \} $
near the horizon (and thus, the $k \to \infty $ behaviour of the
Bogoliubov coefficients )
is enough to determine the relevant quantities
of interest. The expectation value of the Schwarzschild number
operator in the Kruskal vacuum is determined by the function
$$
N(E,E') ~ \equiv ~ <0_K | \a_E^{\ast} \;  \a_{E'} |0_K > ~  =
\int_{-\infty}^{\infty} dk ~ {\cal B}_{kE}^{\ast} ~ {\cal B}_{kE'}
\eqn\prod
$$
The $E \to E'$ limit is determined by the $k \to \infty$ behavior
of eq.\prod , namely
$$
N(E,E') = N_V(E) \; \delta (E - E') ~~,~~ N_V(E) \; = \; { 1 \o {e^{2\pi E} -
1}}
$$
$N_V(E)$ is the number of created modes per unit frequency and per unit
volume of the Schwarzschild manifold, at temperature $T = 1/(2\pi)$.

This can be seen directly from the mapping eq.\restb\ $u = f(u')$ , i.e.,
$$
u ~ = ~ e^{ r^{\star} + t} ~~,~~ v ~ = ~ -  e^{ r^{\star} - t}~.
$$
which defines the maximal extension of the manifold [\ns ]. This mapping is
holomorphic in the complex variable $ r^{\star} - i\Theta $ and has a
critical point defining the event horizon, i.e. $f^{\prime}(-\infty) =
0$.
The inverse of the periodicity in the imaginary  time $\Theta$ is the
temperature characterizing the vacuum spectrum $N_V(\lambda)$.

A complete classification of the vacuum spectra and thermal properties
in a large class of curved and accelerated manifolds can be done
directly in terms of the properties of the holomorphic mappings
  $u = f(u^\prime)$
defining the maximal analytic extension of the manifold, of which
the exponential mapping eq.\restb\
is just a particular case, ($u$ being global or Kruskal-like
coordinates, $u^{\prime}$ being accelerated or Schwarzschild-like
ones) [\ns].

{\bf ACKNOWLEDGMENTS}

 The authors acknowledge the french-spanish scientific
 cooperation for partial support sponsored by Ministerio de
Educaci\'on y Ciencia (DGICYT) (Spain), Minist\`ere des
Affaires Etrang\`eres and Minist\`ere de L'Education Nationale
(France).

\refout

\bye